\documentclass{pasj01}
\usepackage{natbib}
\usepackage{lineno}

\newcommand{\revA}{}
\newcommand{\revB}{}
\newcommand{\revC}{}
\newcommand{\revD}{}
\newcommand{\revE}{}

\title{\LETTERLABEL TOI-2285b: A 1.7 Earth-radius Planet Near the Habitable Zone around a Nearby M Dwarf}

\begin{document}

\author{
Akihiko~\textsc{Fukui},\altaffilmark{1,2}
Tadahiro~\textsc{Kimura},\altaffilmark{3}
Teruyuki~\textsc{Hirano},\altaffilmark{4}
Norio~\textsc{Narita},\altaffilmark{1,5,4,2}
Takanori~\textsc{Kodama},\altaffilmark{1}
Yasunori~\textsc{Hori},\altaffilmark{4,6}
Masahiro~\textsc{Ikoma},\altaffilmark{6}
Enric~\textsc{Pall\'e},\altaffilmark{2,7}
Felipe~\textsc{Murgas},\altaffilmark{2,7}
Hannu~\textsc{Parviainen},\altaffilmark{2,7}
Kiyoe~\textsc{Kawauchi},\altaffilmark{2}
Mayuko~\textsc{Mori},\altaffilmark{8}
Emma~\textsc{Esparza-Borges},\altaffilmark{2,7}
Allyson~\textsc{Bieryla},\altaffilmark{9}
Jonathan~\textsc{Irwin},\altaffilmark{9}
Boris~S.~\textsc{Safonov},\altaffilmark{10}
Keivan~G.~\textsc{Stassun},\altaffilmark{11,12}
Leticia~\textsc{Alvarez-Hernandez},\altaffilmark{7}
V\'ictor~J.~S.~\textsc{B\'ejar},\altaffilmark{2,7}
N\'uria~\textsc{Casasayas-Barris},\altaffilmark{13}
Guo~\textsc{Chen},\altaffilmark{14}
Nicolas~\textsc{Crouzet},\altaffilmark{15}
Jerome~P.~\textsc{de Leon},\altaffilmark{8}
Keisuke~\textsc{Isogai},\altaffilmark{16,17}
Taiki~\textsc{Kagetani},\altaffilmark{17}
Peter~\textsc{Klagyivik},\altaffilmark{18}
Judith~\textsc{Korth},\altaffilmark{19}
Seiya~\textsc{Kurita},\altaffilmark{3}
Nobuhiko~\textsc{Kusakabe},\altaffilmark{4}
John~\textsc{Livingston},\altaffilmark{8}
Rafael~\textsc{Luque},\altaffilmark{20}
Alberto~\textsc{Madrigal-Aguado},\altaffilmark{2,7}
Giuseppe~\textsc{Morello},\altaffilmark{2,7}
Taku~\textsc{Nishiumi},\altaffilmark{21,4}
Jaume~\textsc{Orell-Miquel},\altaffilmark{2,7}
Mahmoudreza~\textsc{Oshagh},\altaffilmark{2,7}
Manuel~\textsc{S\'anchez-Benavente},\altaffilmark{2,7}
Monika~\textsc{Stangret},\altaffilmark{2,7}
Yuka~\textsc{Terada},\altaffilmark{22,23}
Noriharu~\textsc{Watanabe},\altaffilmark{17}
Yujie~\textsc{Zou},\altaffilmark{17}
Motohide~\textsc{Tamura},\altaffilmark{8,4,6}
Takashi~\textsc{Kurokawa},\altaffilmark{4,24}
Masayuki~\textsc{Kuzuhara},\altaffilmark{4,6}
Jun~\textsc{Nishikawa},\altaffilmark{6,21,4}
Masashi~\textsc{Omiya},\altaffilmark{4,6}
S\'ebastien~\textsc{Vievard},\altaffilmark{25}
Akitoshi~\textsc{Ueda},\altaffilmark{4,6,21}
David~W.~\textsc{Latham},\altaffilmark{9}
Samuel~N.~\textsc{Quinn},\altaffilmark{9}
Ivan~S.~\textsc{Strakhov},\altaffilmark{10}
Alexandr~A.~\textsc{Belinski},\altaffilmark{10}
Jon~M.~\textsc{Jenkins},\altaffilmark{26}
George~R.~\textsc{Ricker},\altaffilmark{27}
Sara~\textsc{Seager},\altaffilmark{27,28,29}
Roland~\textsc{Vanderspek},\altaffilmark{27}
Joshua~N.~\textsc{Winn},\altaffilmark{30}
David~\textsc{Charbonneau},\altaffilmark{9}
David~R.~\textsc{Ciardi},\altaffilmark{31}
Karen~A.~\textsc{Collins},\altaffilmark{9}
John~P.~\textsc{Doty},\altaffilmark{32}
Etienne~\textsc{Bachelet},\altaffilmark{33}
Daniel~\textsc{Harbeck},\altaffilmark{33}
}

\altaffiltext{1}{Komaba Institute for Science, The University of Tokyo, 3-8-1 Komaba, Meguro, Tokyo 153-8902, Japan}
\altaffiltext{2}{Instituto de Astrof\'isica de Canarias, V\'ia L\'actea s/n, E-38205 La Laguna, Tenerife, Spain}
\altaffiltext{3}{Department of Earth and Planetary Science, Graduate School of Science, The University of Tokyo, 7-3-1 Hongo, Bunkyo-ku, Tokyo 113-0033, Japan}
\altaffiltext{4}{Astrobiology Center, 2-21-1 Osawa, Mitaka, Tokyo 181-8588, Japan}
\altaffiltext{5}{Japan Science and Technology Agency, PRESTO, 3-8-1 Komaba, Meguro, Tokyo 153-8902, Japan}
\altaffiltext{6}{National Astronomical Observatory of Japan, 2-21-1, Osawa, Mitaka, 181-8588 Tokyo, Japan}
\altaffiltext{7}{Departamento de Astrof\'isica, Universidad de La Laguna, 38206 La Laguna, Tenerife, Spain}
\altaffiltext{8}{Department of Astronomy, Graduate School of Science, The University of Tokyo, 7-3-1 Hongo, Bunkyo-ku, Tokyo 113-0033, Japan}
\altaffiltext{9}{Center for Astrophysics \textbar \ Harvard \& Smithsonian, 60 Garden Street, Cambridge, MA 02138, USA}
\altaffiltext{10}{Sternberg Astronomical Institute, M.V. Lomonosov Moscow State University, 13, Universitetskij pr., 119234, Moscow, Russia}
\altaffiltext{11}{Department of Physics and Astronomy, Vanderbilt University, 6301 Stevenson Center Ln., Nashville, TN 37235, USA}
\altaffiltext{12}{Department of Physics, Fisk University, 1000 17th Avenue North, Nashville, TN 37208, USA}
\altaffiltext{13}{Leiden Observatory, Leiden University, Postbus 9513, 2300 RA Leiden, The Netherlands}
\altaffiltext{14}{CAS Key Laboratory of Planetary Sciences, Purple Mountain Observatory, Chinese Academy of Sciences, Nanjing, 210023, PR China}
\altaffiltext{15}{European Space Agency (ESA), European Space Research and Technology Centre (ESTEC), Keplerlaan 1, 2201 AZ Noordwijk, The Netherlands}
\altaffiltext{16}{Okayama Observatory, Kyoto University, 3037-5 Honjo, Kamogatacho, Asakuchi, Okayama 719-0232, Japan}
\altaffiltext{17}{Department of Multi-Disciplinary Sciences, Graduate School of Arts and Sciences, The University of Tokyo, 3-8-1 Komaba, Meguro, Tokyo 153-8902, Japan}
\altaffiltext{18}{Institute of Planetary Research, German Aerospace Center, Rutherfordstrasse 2, 12489, Berlin, Germany}
\altaffiltext{19}{Department of Space, Earth and Environment, Astronomy and Plasma Physics, Chalmers University of Technology, 412 96 Gothenburg, Sweden}
\altaffiltext{20}{Instituto de Astrof\'isica de Andaluc\'ia (IAA-CSIC), Glorieta de la Astronom\'ia s/n, 18008 Granada, Spain}
\altaffiltext{21}{Department of Astronomical Science, The Graduated University for Advanced Studies, SOKENDAI, 2-21-1, Osawa, Mitaka, Tokyo, 181-8588, Japan}
\altaffiltext{22}{Institute of Astronomy and Astrophysics, Academia Sinica, P.O. Box 23-141, Taipei 10617, Taiwan, R.O.C.}
\altaffiltext{23}{Department of Astrophysics, National Taiwan University, Taipei 10617, Taiwan, R.O.C.}
\altaffiltext{24}{Tokyo University of Agriculture and Technology, 2-24-16, Naka-cho, Koganei, Tokyo, 184-8588, Japan}
\altaffiltext{25}{Subaru Telescope, 650 N. Aohoku Place, Hilo, HI 96720, USA}
\altaffiltext{26}{NASA Ames Research Center, Moffett Field, CA  94035, USA}
\altaffiltext{27}{Department of Physics and Kavli Institute for Astrophysics and Space Research, Massachusetts Institute of Technology, 77 Massachusetts Avenue, Cambridge, MA 02139, USA}
\altaffiltext{28}{Department of Earth, Atmospheric and Planetary Sciences, Massachusetts Institute of Technology, 77 Massachusetts Avenue, Cambridge, MA 02139, USA}
\altaffiltext{29}{Department of Aeronautics and Astronautics, Massachusetts Institute of Technology, 77 Massachusetts Avenue, Cambridge, MA 02139, USA}
\altaffiltext{30}{Department of Astrophysical Sciences, Princeton University, 4 Ivy Lane, Princeton, NJ 08540, USA}
\altaffiltext{31}{Caltech/IPAC-NASA Exoplanet Science Institute, 770 S. Wilson Avenue, Pasadena, CA 91106, USA}
\altaffiltext{32}{Noqsi Aerospace Ltd., 15 Blanchard Avenue, Billerica, MA 01821, USA}
\altaffiltext{33}{Las Cumbres Observatory, 6740 Cortona Drive, Suite 102, Goleta, CA 93117-5575, USA}

\email{afukui@g.ecc.u-tokyo.ac.jp}


\maketitle

\begin{abstract}

We report the discovery of TOI-2285b, a sub-Neptune-sized planet transiting a nearby (42~pc) M dwarf with a period of 27.3 days. We identified the transit signal from the \revE{Transiting Exoplanet Survey Satellite} photometric data, which we confirmed with ground-based photometric observations using the multiband imagers MuSCAT2 and MuSCAT3. Combining these data with other follow-up observations including high resolution spectroscopy with \revE{the Tillinghast Reflector Echelle Spectrograph}, high resolution imaging with \revE{the SPeckle Polarimeter}, and radial velocity (RV) measurements with \revE{the InfraRed Doppler instrument}, we find that the planet has a radius of $1.74 \pm 0.08$~$R_\oplus$, a mass of $<$~\revC{19.5}~$M_\oplus$ (95\% c.l.), and an insolation flux of $1.54 \pm 0.14$ times that of the Earth. Although the planet resides just outside the habitable zone for a rocky planet, if the planet harbors an H$_2$O layer under a hydrogen-rich atmosphere, then liquid water could exist on the surface of the H$_2$O layer depending on the planetary mass and water mass fraction. The bright host star in \revE{the} near infrared ($K_s=9.0$) makes this planet an excellent target for further RV and atmospheric observations to improve our understanding on the composition, formation, and habitability of \revC{sub-Neptune-sized planets}.

\end{abstract}


\section{Introduction}

The Kepler space mission has revealed that planets with sizes between the Earth and Neptune (hereafter \revE{``sub-Neptune-sized planets''}) are abundant in close-in orbits around stars other than the Sun \citep[e.g.,][]{2011ApJ...736...19B}.
Precise radius measurements for these planets have found that the ``hot'' ($S \gtrsim 10$~ $S_\oplus$, where $S$ is insolation flux) \revC{sub-Neptune-sized planets} are classified into two populations; one is \revE{of} hotter and smaller planets and the other is \revE{of} cooler and larger planets \citep[e.g.,][]{2017AJ....154..109F}, which are often referred to as super-Earths and mini-Neptunes, respectively.
Atmospheric evolution models predict that the super-Earths have rocky compositions, where any hydrogen-rich atmosphere has been stripped away by the photoevaporation and/or core-powered mass loss mechanisms, while the mini-Neptunes still retain primordial hydrogen atmospheres \citep[e.g.,][]{2017ApJ...847...29O,2018MNRAS.476..759G}. This scenario has been supported by mass and/or atmospheric measurements for a subset of these systems \citep[e.g.,][]{2013ApJ...776....2L,2015Natur.522..459E}.
On the other hand, less has been known about the compositions of \revE{``cooler''} \revC{sub-Neptune-sized planets} ($S \lesssim 10$~$S_\oplus$) due to the smaller number of discoveries, in particular around bright host stars that allow for various follow-up observations including mass measurements and atmospheric observations. Increasing the sample of such planets is important to construct a comprehensive picture of the compositions and formation histories of \revC{sub-Neptune-sized planets}.

Cooler \revC{sub-Neptune-sized planets} also have the exciting possibility that they may retain liquid water under a hydrogen atmosphere, even if they reside outside of the habitable zone \revA{for rocky planets} \citep{2021MNRAS.505.3414N,2021arXiv210810888M}. If the planets have an H$_2$O layer beneath the hydrogen atmosphere like Neptune and Uranus, then the hydrogen-H$_2$O boundary could have the right conditions for H$_2$O to be liquid.
Although it is not clear if life can exist on such planets, because their atmospheres are easier to observe compared to Earth-like planets thanks to the larger planetary size and larger atmospheric scale height (lower mean-molecular weight), they could potentially be realistic targets for biomarker searches in the next decades \citep{2013ApJ...777...95S,2021arXiv210810888M}.

Here we report the discovery of a new temperate sub-Neptune transiting a nearby M dwarf from the \revE{Transiting
Exoplanet Survey Satellite (TESS)} photometric survey \citep{2015JATIS...1a4003R} and ground-based follow-up observations.

\section{Observations}
\label{sec:obs}

\subsection{TESS photometry}

\begin{figure*}[t]
    \begin{center}
    \includegraphics[width=17cm]{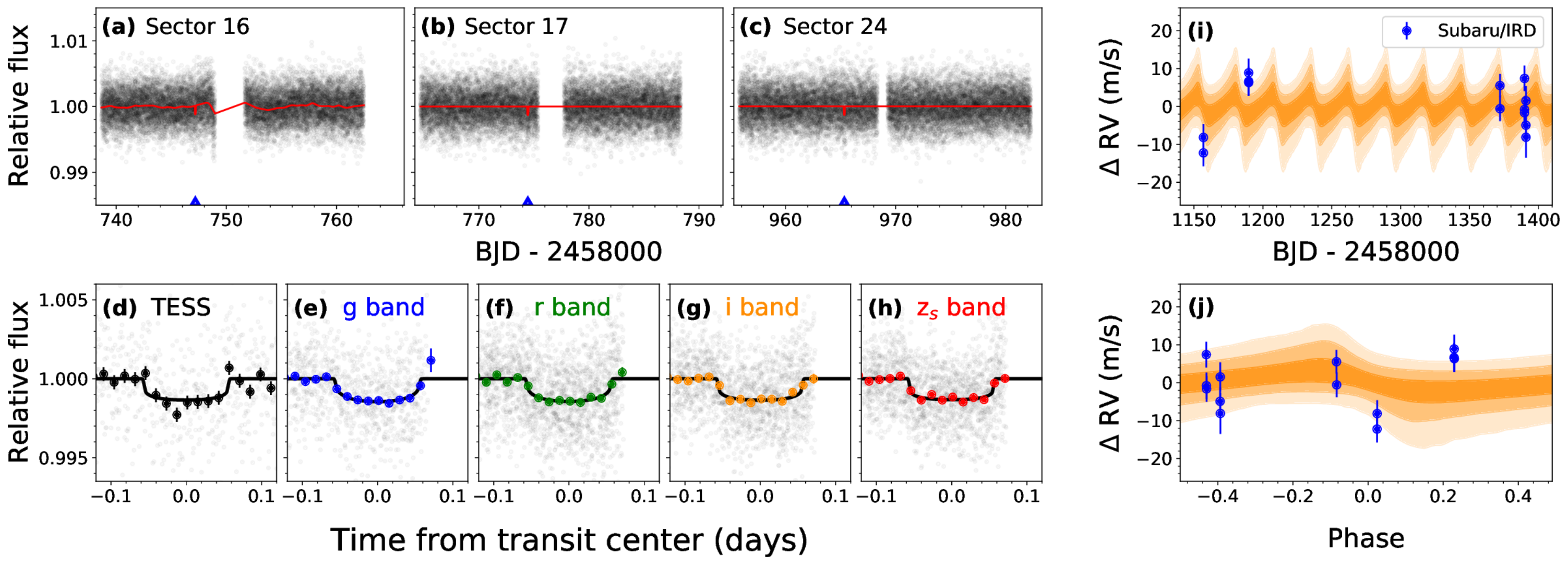}
    \end{center}
    \caption{(a)--(c) PDC-SAP light curves of TOI-2285 from TESS Sectors 16, 17, and 24, respectively. The red lines show best-fit transit+systematic models, and blue triangles indicate the locations of transits of TOI-2285b. (d) Systematic-corrected and phase-folded transit light curves from TESS. The gray dots and filled circles represent individual exposure data and 20-minute-binned data, respectively. (e)--(h) Same as (d) but from the ground (MuSCAT2 and MuSCAT3). \revC{(i) Relative radial velocity of TOI-2285 as a function of time measured with Subaru/IRD (blue points). The orange shades indicate 1$\sigma$, 2$\sigma$, and 3$\sigma$ confidence regions from dark to light, respectively. (j) Same as (i), but phase-folded. } (Color online)}
    \label{fig:lc}
\end{figure*}

TOI-2285 (TIC~329148988) \revB{is an M~dwarf located at a distance of 42~pc \citep{Stassun_2019} with astrometric properties and magnitudes listed in table~\ref{tbl:stellar_properties}. This star} was observed by TESS with 2~min cadences in Sectors 16, 17, and 24, each of which lasted for 25--27~days between 2019 September 12 \revB{and} 2020 May 12 UT. 
The collected data were processed with a pipeline developed by the TESS Science Processing Operations Center (SPOC) at NASA Ames Research Center \citep{2016SPIE.9913E..3EJ}, from which a transit signal with an orbital period of 27.270~d was identified using dedicated pipelines \citep{2010SPIE.7740E..0DJ,Twicken:DVdiagnostics2018}. This planetary candidate was released as TOI-2285.01 (hereafter TOI-2285b) on 2020 September 30 UT by TESS Science Office at \revE{Massachusetts Institute of
Technology} \citep{2021arXiv210312538G}. We \revC{downloaded} the Presearch Data Conditioning Simple Aperture Photometry (PDC-SAP) \citep[][and references therein]{2014PASP..126..100S}
from the Mikulski Archive for Space Telescopes (MAST) at the Space Telescope Science Institute. \revC{The normalized PDC-SAP light curves are} shown in figure~\ref{fig:lc}.

\begin{table}[h]
\caption{Properties of the host star TOI-2285.}
\label{tbl:stellar_properties}
\begin{center}
\begin{tabular}{lcc}
\hline
Parameter & Value & Reference\footnotemark[$*$]\\
\hline
\multicolumn{3}{c}{\it{Astrometric and kinematic parameters}}\\
$\alpha$ \revA{(epoch J2016.0)} & 22:10:15.185 & (1)\\
$\delta$ \revA{(epoch J2016.0)} & +58:42:21.93 & (1)\\
$\mu_{\alpha}$\,cos\,$\delta$ (mas\,yr$^{-1}$)  & $21.263 \pm 0.054$ & (1)\\
$\mu_{\delta}$ (mas\,yr$^{-1}$) & $-20.874 \pm 0.044$ & (1)\\
Distance (pc) & $42.409 \pm 0.047$ & (2)\\
RV (km~s$^{-1}$) & $-24.1 \pm 0.5$ & This work\\
$U$ (km s$^{-1}$) & 4.45 $\pm$ 0.12 & This work\\
$V$ (km s$^{-1}$) & -23.48 $\pm$ 0.50 & This work\\
$W$ (km s$^{-1}$) & -6.78 $\pm$ 0.02  & This work\\
\multicolumn{3}{c}{\it{Magnitudes}}\\
$V$ & $13.403 \pm 0.092$ & (2)\\
TESS & $11.3078 \pm 0.0073$ & (2)\\
$J$ & $9.860 \pm 0.027$ & (3)\\
$H$ & $9.262 \pm 0.028$ & (3)\\
$K_s$ & $9.034 \pm 0.022$ & (3)\\
\multicolumn{3}{c}{\it{Physical parameters}}\\
Mass ($M_\odot$) & $0.454 \pm 0.010$ & This work\\
Radius ($R_\odot$) & $0.464 \pm 0.013$ & This work\\
Luminosity ($L_\odot$) & $0.0287 \pm 0.0010$ & This work\\
$T_{\rm eff}$ (K) & $3491 \pm 58$ & This work\\
$[$Fe/H$]$ (dex) & $-0.05 \pm 0.12$ & This work\\
\hline
\end{tabular}
\end{center}
\begin{tabnote}
\footnotemark[$*$] References: (1) Gaia EDR3 \citep{2021A&A...649A...1G};  (2) TIC v8 \citep{Stassun_2019}; (3) 2MASS \citep{2006AJ....131.1163S}
\end{tabnote}
\end{table}

\subsection{Speckle imaging with SAI 2.5m/SPP}

TOI-2285 was observed on 2020 December 21 UT with the SPeckle Polarimeter \citep[SPP;][]{Safonov+2017} on the 2.5~m telescope at the Caucasian Observatory of Sternberg Astronomical Institute (SAI) of Lomonosov Moscow State University. SPP uses Electron Multiplying CCD Andor iXon 897 as a detector. The atmospheric dispersion compensator allowed observation of this relatively faint target through the wide-band $I_c$ filter. The power spectrum was estimated from 4000 frames with 30 ms exposure. The detector has a pixel scale of 20.6 mas pixel$^{-1}$, and the angular resolution was 89 mas. We did not detect any stellar companions brighter than $\Delta {\rm mag}=$2.7 and 4.2 at 0\farcs25 and 0\farcs5, respectively.

\subsection{High-resolution spectroscopy with TRES}

We obtained reconnaissance spectra of TOI-2285 on 2020 October 11 and 25 UT using the Tillinghast Reflector Echelle Spectrograph \citep[TRES;][]{2008psa..conf..287F} on the 1.5~m Tillinghast telescope at the Fred Lawrence Whipple Observatory (FLWO) in Arizona. TRES has a resolving power of $\approx 44,000$ and a wavelength coverage of 385--910~nm. 
The spectra were extracted as described in \citet{2010ApJ...720.1118B}. No rotational broadening was detected in the spectra.
The systemic radial velocity (RV) was derived to be $-24.1 \pm 0.5$~km~s$^{-1}$ following methods described by \citet{2018AJ....155..125W} using cross correlation against an observed template spectrum of Barnard's Star, for which a barycentric RV of $-110.3 \pm 0.5$~km~s$^{-1}$ was adopted.

\subsection{Transit photometry with MuSCATs}

We observed two full transits of TOI-2285b on 2020 October 6 UT and 2021 July 5 UT with the multiband imagers MuSCAT3 \citep{2020SPIE11447E..5KN} on the 2~m \revA{FTN} telescope \revA{of Las Cumbres Observatory} at the Haleakala observatory, Hawaii and MuSCAT2 \citep{2019JATIS...5a5001N} on the 1.52m Telescopio Carlos S\'{a}nchez (TCS) at the Teide observatory, Spain, respectively.
Both instruments have four channels for the $g$, $r$, $i$, and $z_s$ bands, and each channel \revD{of MuSCAT3 (MuSCAT2)} is equipped with a 2k $\times$ 2k \revD{(1k $\times$ 1k)} CCD camera with a pixel scale of 0\farcs266 pixel$^{-1}$ (\revD{0\farcs435 pixel$^{-1}$}).
Both observations were slightly defocused to avoid saturation, with exposure times ranging from 10 to 30~s depending on the band and instrument.
After calibrating the obtained images for dark and flat fields, we extracted light curves by aperture photometry using a custom pipeline \citep{2011PASJ...63..287F} with aperture radii of 4$^{\prime\prime}$--6$^{\prime\prime}$. The resultant photometric dispersion per exposure ranges from 0.22\% to 0.32\% depending on the band and instrument.

\subsection{Radial velocity measurements with Subaru/IRD}

We obtained high-resolution spectra of TOI-2285 using the InfraRed Doppler (IRD) instrument on the 8.2m Subaru telescope \citep{2012SPIE.8446E..1TT, 2018SPIE10702E..11K} on seven nights between 2020 October 30 to 2021 June 25 UT, under the Subaru-IRD TESS intensive follow-up program (ID: S20B-088I). IRD is a fiber-fed spectrograph covering the near infrared wavelengths from 930 nm to 1740 nm with a spectral resolution of $\approx 70,000$. 
The integration time per exposure was set to 600--1200~s depending on the observing condition. We also observed at least one telluric standard star (A0 or A1 star) on each night to correct for the telluric lines in extracting the template spectrum for the RV analysis. 
Raw IRD data were reduced by the procedure described in \citet{2020PASJ...72...93H}, where wavelengths were calibrated by spectra of a laser-frequency comb. The reduced one-dimensional spectra have a typical signal-to-noise ratio (SNR) of 50--90 per pixel at 1000 nm. \revD{W}e discarded the data with very low SNR ($\sim$20), which could be affected by detector persistence. 
We extracted RV for each frame following the procedure of \citet{2020PASJ...72...93H}\revC{, as shown in panel (i) of figure~\ref{fig:lc}}.
The typical RV internal errors are 3--4~m~s$^{-1}$.

\section{Analysis and Results}
\label{sec:analysis}

\subsection{Stellar properties}
The physical parameters of the host star were derived as follows.
First, from the TRES spectra, we empirically derived the stellar radius and iron abundance to be $R_s = 0.453 \pm 0.045$~$R_\odot$ and [Fe/H]$=-0.05 \pm 0.12$~dex, respectively, using {\tt SpecMatch-Emp} \citep{2017ApJ...836...77Y}. Next, we performed an analysis of the broadband spectral energy distribution (SED) of the star, together with the Gaia EDR3 parallax and the [Fe/H] value derived above, following \cite{Stassun:2017}.
We estimated the stellar bolometric luminosity and effective temperature to be $L_{\rm bol} = 0.0287 \pm 0.0010$~$L_\odot$ and $T_{\rm eff}=3450 \pm 50$~K, respectively. These two parameters provide an independent estimate on $R_s$ of $0.475 \pm 0.016$~$R_\odot$ via the Stefan-Boltzmann law. We also independently obtained $R_s = 0.458 \pm 0.013$~$R_\odot$ using the empirical absolute-$K_s$-metallicity-radius relation of \citet{2015ApJ...804...64M}.
We took a weighted mean of the above three estimations of $R_s$ for its final value, while conservatively adopting the maximum uncertainty among the three for the uncertainty of the final value \revD{taking into account possible systematic errors, leading to
$R_s = 0.464 \pm 0.013$~$R_\odot$.}
The $T_{\rm eff}$ value was then self-consistently updated to be $3491 \pm 58$~K using $L_{\rm bol}$ and the final $R_s$ value.
Finally, we determined the stellar mass to be $M_s = 0.454 \pm 0.010$~$M_\odot$ using the empirical absolute-$K_s$-metallicity-mass relation of \citet{2019ApJ...871...63M}.

We also investigated the activity and age of the host star. We found an absorption (not emission) line of $H\alpha$ in the TRES spectrum with an equivalent width of $0.180 \pm 0.027$~\AA, indicating that the star is inactive and not young. We searched for photometric variabilities originated from stellar rotation in the TESS data, ASAS-SN public light curves\footnote{https://asas-sn.osu.edu/}, and ZTF public light curves\footnote{https://irsa.ipac.caltech.edu/Missions/ztf.html}, but found no significant periodic variability in any of these data, confirming the inactiveness of the star. The sky coordinates and proper motion of the star do not match with any stellar moving groups. On the other hand, the Galactic space velocities of the star calculated from the Gaia's astrometric measurements and systemic RV derived by the TRES spectra, as listed in table~\ref{tbl:stellar_properties}, are consistent with a member of the Galactic thin disk.
In summary, the host star is not a very young or old, but is a typical field star with no significant chromospheric activity.

\begin{table}
\caption{Planetary parameters.}
\label{tbl:planetary_parameters}
\begin{center}
\begin{tabular}{lr}
\hline
Parameter & Value\\
\hline
\multicolumn{2}{c}{\revC{{\it Fitted parameters}}}\\
Orbital period, $P$ (d) & $27.26955 \revC{^{+0.00013}_{-0.00010}}$ \\
Transit epoch, $t_0$ (BJD) &  $2458747.1815 \revC{^{+0.0017}_{-0.0021}}$\\
Impact parameter, $b$  & \revC{$0.49\ ^{+0.19}_{-0.31}$} \\
Radius ratio, $R_p/R_s$ (\%) & $3.44 \pm \revC{0.13}$ \\
\revC{Eccentricity parameter, $\sqrt{e} \cos \omega$} & $\revC{0.24\ ^{+0.34}_{-0.62}}$\\
\revC{Eccentricity parameter, $\sqrt{e} \sin \omega$} & $\revC{0.32\ ^{+0.14}_{-0.18}}$\\
RV semi-amplitude, $K$ (m s$^{-1}$) & $<$\revC{7.5}\footnotemark[$\dag$]\\
RV jitter, $\sigma_{\rm jit}$ (m s$^{-1}$) & $\revC{6.5}\ ^{+2.4}_{-1.8}$\\
\multicolumn{2}{c}{\revC{{\it Derived parameters}}}\\
Scaled semi-major axis, $a/R_s$ & \revC{$63.0\ ^{+1.9}_{-1.7}$} \\
Orbital inclination, $i$ ($^\circ$)  & \revC{$89.66\ ^{+0.22}_{-0.19}$}\\ 
\revC{Eccentricity, $e$} & \revC{$0.30\ ^{+0.10}_{-0.09}$}\\
\revC{Arg. of periastron, $\omega$ ($^\circ$)} & \revC{$56\ ^{+80}_{-39}$}\\
Radius, $R_p$ ($R_\oplus$)  & $1.74 \pm 0.08$\\
Mass, $M_p$ ($M_\oplus$)  & $<$\revC{19.5}\footnotemark[$\dag$]\\
Semi-major axis, $a$ (au)  & $0.1363 \pm 0.0010$\\
Insolation flux, $S$ ($S_\oplus$)  & $1.54 \pm 0.14$\\
Equilibrium temp.\footnotemark[$*$], $T_{\rm eq}$ (K) & $284 \pm 6$
\\
\hline
\end{tabular}
\end{center}
\begin{tabnote}
\footnotemark[$*$] Bond albedo of 0.3 and uniform surface temperature are assumed.\\
\footnotemark[$\dag$] \revA{The value indicates a 95\% confidence upper limit.}
\end{tabnote}
\end{table}


\subsection{\revC{Planet validation and  light-curve + RV joint modeling}}
\label{sec:validation}

We found no significant variations in the IRD data, excluding a false positive scenario that the transiting object is a stellar companion to TOI-2285. 
\revD{We also measured the full-to-total duration ratio of the transit signal to be $0.913\ ^{+0.018}_{-0.036}$, which sets an upper limit of 4.5\% (4$\sigma$) on the true eclipse depth  \citep{2003ApJ...585.1038S}. This, combined with the SPP speckle observation, excludes any eclipsing binary outside of 0\farcs5 from TOI-2285 as the source of the transit signal. The remaining false positive scenario is that the transit signal is caused by an eclipsing binary within 0\farcs5, the probability of which is calculated to be only $3.6 \times 10^{-4}$ by {\tt vespa} \citep{Morton2015}.}

\revC{To derive the physical parameters of the \revD{now-validated} planet, we fit a transit+RV model simultaneously to the transit light curves and RV data. For the transit model, we used a Mandel--Agol model implemented by {\tt PyTransit} \citep{Parviainen2015}. The fitting parameters include impact parameter $b$, planet-to-star radius ratio $R_p/R_s$, orbital period $P$, reference transit time $t_0$, RV semi-amplitude $K$, RV zero point $V_0$, RV jitter $\sigma_{\rm jit}$, stellar mass $M_s$, stellar radius $R_s$, and two eccentricity components $\sqrt{e} \cos \omega$ and $\sqrt{e} \sin \omega$, where $e$ and $\omega$ are the eccentricity and argument of periastron, respectively. For stellar limb-darkening, we applied a quadratic limb-darkening law with two coefficients $u_1$ and $u_2$, which we fixed to theoretical values calculated by {\tt LDTk} \citep{Parviainen2015b} for a star with parameters listed in table \ref{tbl:stellar_properties}.}

Simultaneously with the transit model, we also modeled time-correlated noise in \revC{the} TESS and ground-based data following the procedure described in \citet{2021arXiv210705430F}.
In short, time-correlated noise in the TESS light curve was modeled by a Gaussian process (GP) implemented in {\tt celerite} \citep{2017AJ....154..220F} with a kernel function of stochastically-driven, damped simple harmonic oscillator (SHO), where the model parameters are the frequency of undamped oscillation $\omega_0$, the scale factor to the amplitude of the kernel function $S_0$, and quality factor $\mathcal{Q}$. We fit $\omega_0$ and $S_0$ for each sector and fix $\mathcal{Q}$ to unity for all sectors. The time-correlated noise in the ground-based light curves was modeled by a combination of a linear function of the stellar displacements on the detector and a GP model as a function of time with an approximated Mat\'{e}rn 3/2 kernel implemented in {\tt celerite}. The model parameters include two coefficients of the linear function $c_x$ and $c_y$, the signal standard deviation of the GP kernel function $\sigma$, and the length scale of the GP kernel function $\rho$. We \revB{let $c_x$, $c_y$, and $\sigma$ be free for each light curve, while $\rho$ is shared within all bands for each transit.} 

The posterior probabilities of the parameters were sampled by an Markov Chain Monte Carlo (MCMC) method implemented in {\tt emcee} \citep{2013PASP..125..306F}. \revC{We applied uniform priors for all parameters but $M_s$ and $R_s$, for which we applied Gaussian priors with values listed in table~\ref{tbl:stellar_properties}.} 
\revC{Because $e$ and $\omega$ cannot be uniquely constrained due to the lack of planetary signal in the RV data, we restrict the values of $\sqrt{e}\cos \omega$ and $\sqrt{e}\sin \omega$ such that $e < 0.45$, which is an empirical upper limit on the eccentricity of super-Earths and Neptunes around M dwarfs \citep{2011arXiv1109.2497M}.} 
The derived median values and 1$\sigma$ uncertainties are reported in table \ref{tbl:planetary_parameters}, and phase-folded transit light curves \revC{and RV data} are shown in figure~\ref{fig:lc}.
From this analysis, \revC{we derived the planetary radius, eccentricity, and 95\% confidence upper limit on the mass to be $R_p = 1.74 \pm 0.08$~$R_\oplus$, $e = 0.30\ ^{+0.10}_{-0.09}$, and $M_p < 19.5$~$M_\oplus$, respectively.}

\section{Discussion and Summary}
\label{sec:discuss}

\begin{figure}
    \begin{center}
    \includegraphics[height=7.cm]{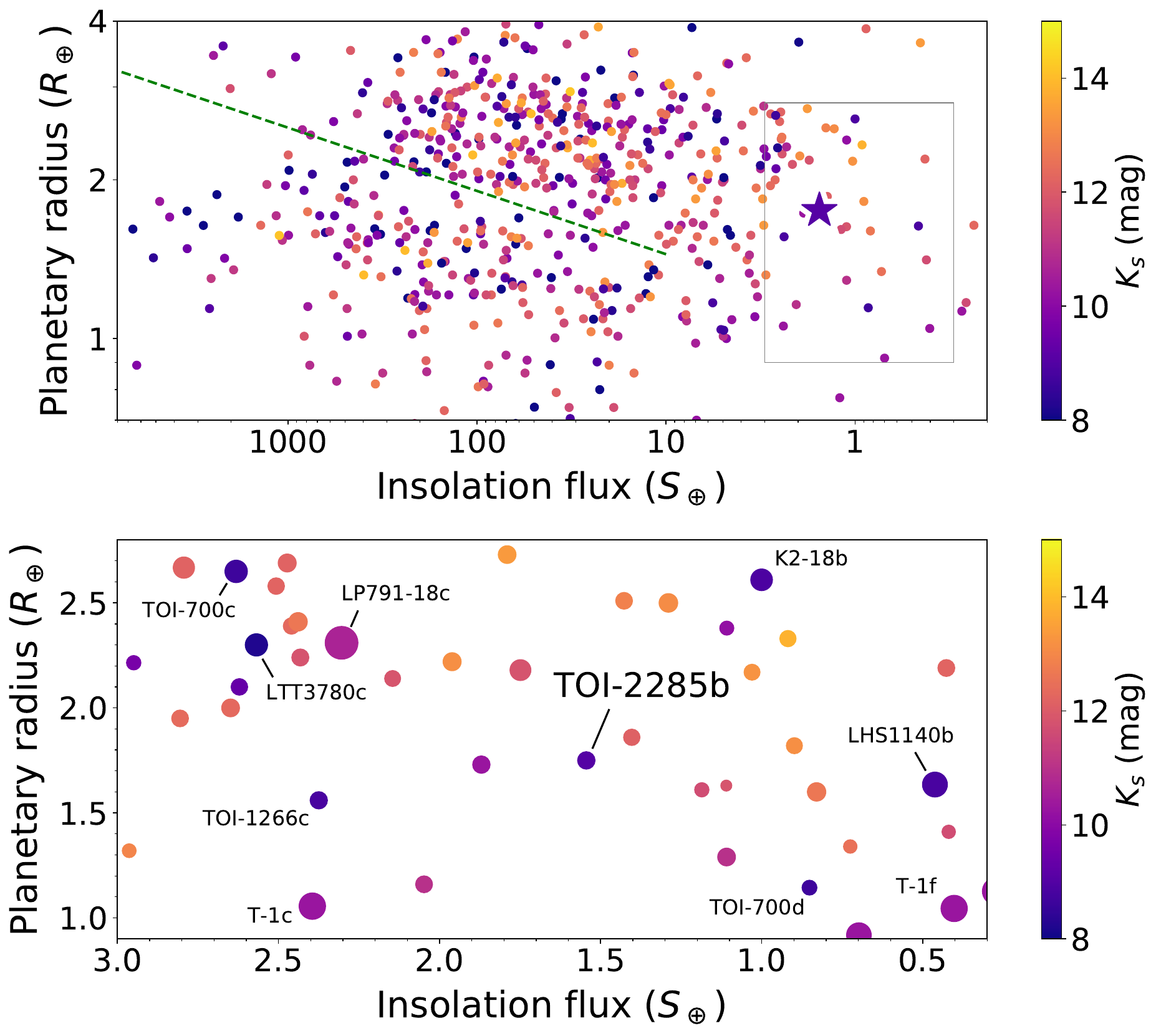}
    \end{center}
    \caption{(Top) Distribution of radius and insolation flux of known transiting planets with radii measured with precision better than 15\%. Colors represent the $K_s$-band brightness of the host star. TOI-2285b is indicated by a star. The green dashed line indicates the location of radius valley for the Kepler's sample proposed by \citet{2019ApJ...875...29M}. (Bottom) Zoom-in of the gray rectangle region in the top panel. The area of each point is scaled by transit depth ($= R_p^2/R_s^2$). ``T-1'' stands for TRAPPIST-1. (Color online)}
    \label{fig:ins_vs_Rp}
\end{figure}

We have \revC{validated} a temperate \revA{1.7~$R_\oplus$ planet} transiting the nearby M dwarf TOI-2285 from the TESS photometric survey \revA{and} ground-based followup observations. Figure \ref{fig:ins_vs_Rp} shows the location of this planet in the insolation-flux vs. planetary-radius plane\footnote{Data in figure~\ref{fig:ins_vs_Rp} are taken from the NASA Exoplanet Archive $<$https://exoplanetarchive.ipac.caltech.edu/$>$.}.
\revA{TOI-2285b is located right in the radius valley found in the Kepler's sample \citep[$\sim$1.5--2.0~$R_\oplus$,][]{2017AJ....154..109F}, while its insolation flux, $1.54 \pm 0.14$~$S_\oplus$, is much lower than the majority of this sample. TOI-2285b is one of only a handful of low-insolation ($S < 3$~$S_\oplus$) \revC{sub-Neptune-sized planets} transiting bright host stars ($K_s < 10$), among which LHS1140b \citep{2017Natur.544..333D} and TOI-1266c \citep{2020A&A...642A..49D} have similar sizes to TOI-2285b (see figure~\ref{fig:ins_vs_Rp}) with measured masses.
Despite their similarity in size and irradiation flux, these two planets probably have different compositions; LHS1140b \citep[$6.5 \pm 0.5$~$M_\oplus$,][]{2020AA...642A.121L} and TOI-1266c \citep[2.2$^{+2.0}_{-1.5}$~$M_\oplus$,][]{2020A&A...642A..49D} are likely rocky and volatile- or gas-rich planets, respectively. 
TOI-2285b will thus be an important sample for a comparative study on the compositions of this class of planets.}

With only the upper limit on the mass ($<$~\revC{19.5}~$M_\oplus$), \revA{however,} there are various possibilities for the bulk composition of this planet. If the true planetary mass is more massive than $\sim8$~$M_\oplus$, then the planet is expected to be a rocky planet without any hydrogen-rich atmosphere, according to the model of \citet{2019PNAS..116.9723Z}. In this case, the planet is probably too hot for water to exist globally as liquid on the surface\revA{,\footnote{Liquid water might still locally exist on the boundary between day-side and night-side; detailed studies of planetary climate with General Circulation Models (GCMs) are required to explore this possibility.} given that the inner edge of the habitable zone for a tidally-locked, rocky planet around a star with $T_{\rm eff} \sim 3500$~K is estimated to be $\sim1.3$~$S_\oplus$ \citep{2018ApJ...852...67H}}.

On the other hand, if the planet is less massive than $\sim$8~$M_\oplus$, then it could have \revA{an H$_2$O layer on top of a rocky core.}
If so, and \revA{the planet is covered by a} 
hydrogen-rich atmosphere, then the \revA{surface of the} H$_2$O layer could be in liquid phase depending on the pressure and temperature at the hydrogen-H$_2$O boundary. To investigate this possibility, 
we numerically integrate a radially-1D hydrostatic-equilibrium structure of the planet that consists of a hydrogen-helium atmosphere on top of a H$_2$O layer on top of a rock central core. 
We find solutions in which the calculated radius at 10 mbar is equal to 1.74~$R_\oplus$. 
We determine the pressure-temperature profile in the same way as \cite{Kurosaki+Ikoma2017} with the radiative equilibrium temperature of 300~K and the intrinsic luminosity of $2\times 10^{20} M_{\rm rock}/M_\oplus$~erg~s$^{-1}$ \citep[see][]{Guillot+1995}. 
The H$_2$O layer and the rocky core are assumed to be fully convective (or iso-entropic). 
We use the tabulated data from \cite{Chabrier+2019} and \cite{Haldemann+2020} for the equations of state (EOSs) of H-He and H$_2$O, respectively. 
Instead of integrating the rocky core structure, we use the data of the mass-radius relationship calculated by \cite{Fortney+2007}. 
The input parameters include the planet’s total mass $M_{\rm p}$, which is defined as the sum of the H$_2$O layer mass ($M_{\rm water}$) and the rocky core mass ($M_{\rm rock}$), and the H$_2$O mass fraction $X_{\rm w}$ (= $M_{\rm water}/M_{\rm p}$). 

\begin{figure}
    \vspace{-20pt}
    \begin{center}
    \includegraphics[width=8cm]{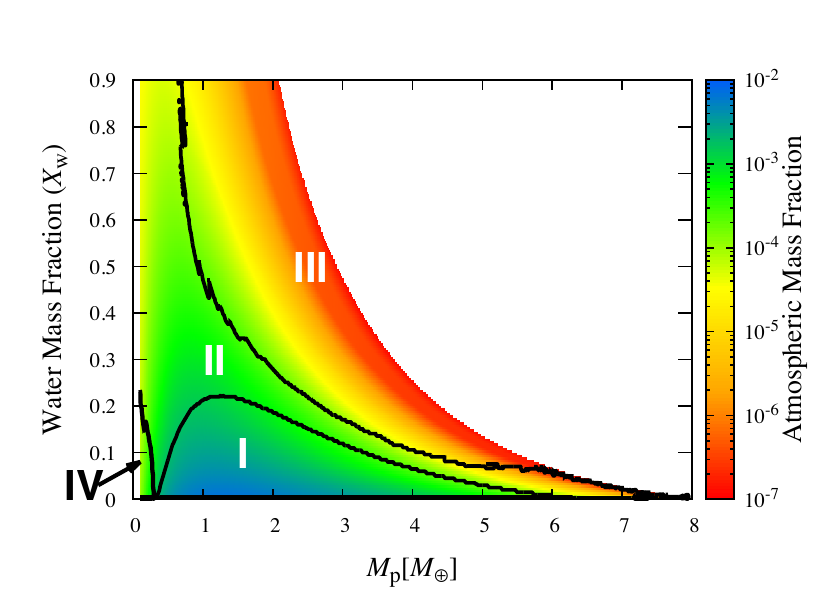}
    \end{center}
    \caption{Contour map of the atmospheric mass fraction as a function of the planet’s total mass ($M_{\rm p}$) and the water-layer mass fraction ($X_{\rm w}$). The solutions are categorized into four regions by black lines according to the water layer states: (I) the entire H$_2$O layer consists of super critical water; (II) a liquid water layer exists on top of a super critical water layer; (III) a liquid water layer exists on top of a high-pressure-ice layer; (IV) the entire H$_2$O layer consists of liquid water, or partially vaporized. There is no solution in the white-colored region. (Color online)}
    \label{fig:Menv_map}
\end{figure}

Figure~\ref{fig:Menv_map} shows the contour map of the atmospheric mass fraction as a function of $M_{\rm p}$ and $X_{\rm w}$.
For a given $X_{\rm w}$, the atmospheric mass is found to decrease with increasing $M_{\rm p}$, because the radius of the H$_2$O layer becomes large, except for a small-$M_{\rm p}$ range ($\lesssim$~1.0~$M_\oplus$) where the gravitational compressibility of the atmosphere is significantly effective.
For a given $M_{\rm p}$, the atmospheric mass is found to decrease with increasing $X_{\rm w}$, because the radius of the H$_2$O layer increases.
The amount of atmospheric gas determines the temperature and pressure at the interface between the atmosphere and the H$_2$O layer, and thus the phase of H$_2$O in the interior.
As indicated in the figure, the solutions are categorized into four regions in the $M_{\rm p}-X_{\rm w}$ plane: (I) the entire H$_2$O layer consists of super critical water; (II) a liquid water layer exists on top of a super critical water layer; (III) a liquid water layer exists on top of a high-pressure-ice layer; (IV) the entire H$_2$O layer consists of liquid water, or partially vaporized. For liquid water to exist in the interior\revA{ (i.e., II--IV)}, the atmospheric mass fraction must be small enough ($\lesssim$~0.5\%). 
\revA{Note that, however, even if the planet still retains such a small amount of hydrogen atmosphere, it could be lost in $\sim$Gyr via photoevaporation \citep[e.g.,][]{2017ApJ...847...29O}.}

To constrain the bulk composition of the planet further, first, it is critical
\revC{to constrain the planetary mass from additional RV observations further. The RV semi-amplitude of TOI-2285 is expected to be $\sim3  \times (M_p/8M_\oplus)$~m~s$^{-1}$, which is within the reach of the current facilities as demonstrated by the achieved RV precision in this work ($\sim$3--4~m~s$^{-1}$).}
Besides, because the mass and radius alone are still not enough to solve completely for the bulk composition, atmospheric investigations through transmission spectroscopy will be key to look into the planetary composition and potential habitability further. \revC{Using Equation (1)  of \cite{2018PASP..130k4401K}, the transit spectroscopic metric (TSM) is estimated to be $23.3  \times  (4 M_\oplus/M_p)$, which makes} such observations feasible with current and upcoming facilities like HST, JWST, and Ariel.

\bigskip

\begin{ack}

Funding for the TESS mission is provided by NASA's Science Mission Directorate.
We acknowledge the use of public TESS data from pipelines at the TESS Science Office and at the TESS SPOC.
Resources supporting this work were provided by the NASA High-End Computing (HEC) Program through the NASA Advanced Supercomputing (NAS) Division at Ames Research Center for the production of the SPOC data products. This paper includes data that are publicly available from the MAST.
This research has made use of the Exoplanet Follow-up Observation Program website, which is operated by the California Institute of Technology, under contract with the NASA under the Exoplanet Exploration Program.
This work makes use of observations from the LCO global telescope network.
This paper is based on observations made with the MuSCAT3 instrument, developed by the Astrobiology Center (ABC) and under financial supports by JSPS KAKENHI (JP18H05439) and JST PRESTO (JPMJPR1775), at FTN on Maui, HI, operated by the LCO, and observations made with the MuSCAT2 instrument, developed by ABC, at TCS operated on the island of Tenerife by the IAC in the Spanish Observatorio del Teide. 
This paper is partially based on observations made at the CMO SAI MSU with the support by M.V. Lomonosov Moscow State University Program of Development.
This work is partly supported by JSPS KAKENHI Grant Numbers 22000005, JP15H02063, JP17H04574, JP18H05439, JP18H05442, JP20K14518, Grant-in-Aid for JSPS Fellows Grant Number JP20J21872, JST PRESTO Grant Number JPMJPR1775, and the ABC of National Institutes of Natural Sciences (Grant Number AB031010).
This work is partly financed by the Spanish Ministry of Economics and Competitiveness through grants PGC2018-098153-B-C31 and PID2019-109522GB-C53.
A.A.B., B.S.S. and I.A.S. acknowledge the support of Ministry of Science and Higher Education of the Russian Federation under the grant 075-15-2020-780 (N13.1902.21.0039). 
N.C.B. and G.M. acknowledge the funding from the European Research Council under the European Union's Horizon 2020 research and innovation program under grant agreement No 694513 and under the Marie Sk\l{}odowska-Curie grant agreement No. 895525, respectively.
J.K. acknowledges the support of the Swedish National Space Agency (SNSA; DNR 2020-00104).
R.L. acknowledges financial support from the Spanish Ministerio de Ciencia e Innovaci\'{o}n, through project PID2019-109522GB-C52/AEI/10.13039/501100011033, and the Centre of Excellence ``Severo Ochoa'' award to the Instituto de Astrof\'{i}sica de Andaluc\'{i}a (SEV-2017-0709).

\end{ack}


\end{document}